\documentclass[aps,pra,twocolumn,superscriptaddress]{revtex4-1}
\usepackage{color}
\usepackage{dcolumn}
\usepackage{graphicx}

\usepackage{amsmath,amsfonts,amssymb}

\usepackage{bbold}
\usepackage{dsfont}

\usepackage{soul}

\newcommand{\oH}{\hat{H}}
\newcommand{\oa}{\hat{a}}
\newcommand{\oad}{\hat{a}^{\dagger}}
\newcommand{\oGamma}{\hat{\Gamma}}
\newcommand{\oGammad}{\hat{\Gamma}^{\dagger}}
\newcommand{\orho}{\hat{\rho}}

\newcommand{\ket}[1]{|#1 \rangle}
\newcommand{\bra}[1]{\langle #1|}
\newcommand{\average}[1]{\langle #1 \rangle}

\begin{document}

\title{Simulating frustrated antiferromagnets with quadratically driven QED cavities}

\author{Riccardo Rota}
\affiliation{Institute of Physics, Ecole Polytechnique F\'ed\'erale de Lausanne (EPFL), CH-1015, Lausanne, Switzerland}
\email{riccardo.rota@epfl.ch}
\author{Vincenzo Savona}
\affiliation{Institute of Physics, Ecole Polytechnique F\'ed\'erale de Lausanne (EPFL), CH-1015, Lausanne, Switzerland}

\date{\today}

\begin{abstract}
	
We propose a class of quantum simulators for antiferromagnetic spin systems, based on coupled photonic cavities in presence of two-photon driving and dissipation. By modeling the coupling between the different cavities through a hopping term with negative amplitude, we solve numerically the quantum master equation governing the dynamics of the open system and determine its non-equilibrium steady state. Under suitable conditions, the steady state can be described in terms of the degenerate ground states of an antiferromagnetic Ising model. When the geometry of the cavity array is incommensurate with the antiferromagnetic coupling, the steady state presents properties which bear full analogy with those typical of the spin liquid phases arising in frustrated magnets.

\end{abstract}

\maketitle
\section{Introduction}

Since many years, quantum simulation has proven very useful to address fundamental problems in different fields of research, from quantum chemistry to condensed-matter physics or cosmology \cite{Lloyd96,Buluta2009,Cirac2012,Johnson2014,Georgescu14,SANCHEZPALENCIA2018,Bloch2018}. Following the pioneering idea of Feynman \cite{Feynman1982}, several experimental platforms have been proposed to implement quantum simulators, neutral atoms in optical lattices \cite{Bloch2012}, trapped ions \cite{Blatt2012}, superconducting circuits \cite{Clarke2008} and photonic systems \cite{Aspuru-Guzik2012}, among others. 

In particular, extended lattices of coupled nonlinear photonic cavities, both at optical and microwave frequencies, has been applied to the simulation of quantum collective phenomena \cite{CiutiRMP,Hartmann16,NohAngelakis16}. The effective photon-photon interaction arising from the nonlinearity of the medium where the electromagnetic field propagates, combined with losses of the cavities, make these systems the ideal platform to investigate the non-equilibrium dynamics of strongly correlated open quantum systems. This has motivated an intense research activity during the last years, which has shown the emergence of interesting phenomena, such as fractional quantum Hall effects \cite{Cho08,Nunnenkamp2011,Umucalilar12,Petrescu12,Hayward12,Roushan2016} or dissipative phase transitions \cite{DallaTorre10,DallaTorre12,Lee13,Sieberer13,Sieberer14,Altman15,Carmichael15,Bartolo16,Mendoza16,CasteelsStorme16,Jin16,Maghrebi16,Marino16,Rota17,Savona17,Casteels17,CasteelsFazio17,FossFeig17,Biondi17,Biella17,Vicentini18,Rota18,Rota19}

A fundamental issue in many-body physics, that is still object of intense investigation, concerns the behavior of frustrated systems. Frustration refers to the presence of competing constraints in the Hamiltonian, which cannot be satisfied simultaneously. This phenomenon is particularly relevant in magnetic systems, where frustration usually has a geometric origin and leads to a macroscopic degeneracy of the ground state \cite{Ramirez94,Moessner06}. Frustrated magnets can be therefore characterized by strong fluctuations even in the limit of zero temperature and display configurations called \textit{spin liquids}, i.e. highly correlated phases with extensive entropy and without static order \cite{Balents2010}. Depending on the nature of the fluctuations, spin liquids can be either classical or quantum. In particular, the latter are prototypical examples of system with long-range entanglement and, although they are more elusive than their classical counterpart, they can show remarkable collective phenomena, such as emergent gauge fields and fractional particle excitations.

Quantum simulators, such as trapped ions \cite{Kim2010} and Rydberg atoms \cite{Glaetzle15,Lienhard18}, have been applied to the study of frustrated magnets. However, although photonic lattices in presence of frustration have been investigated in the past \cite{Masumoto2012,Petrescu12,Jacqmin14,Biondi15,Baboux16,CasteelsRota16,Schmidt_2016,Rota2017Lieb,Biondi18,Arevalo18}, the possibility of simulating spin liquids by means of photonic systems is yet to be explored. An interesting experimental platform able to mimic the behavior of spin systems is represented by QED cavities subjected to two-photon (i.e. quadratic in the field) driving and losses \cite{Leghtas15}. In particular, the two-photon driving scheme enforces a $\mathbb{Z}_2$ symmetry, as it sets the complex phase of the square of the cavity field. This results in the emergence of universal properties characteristic of quantum spin-$1/2$ systems, making thus quadratically-driven photonic resonators a suitable simulator of quantum magnets. Indeed, the non-equilibrium steady state of these systems is approximately restricted to the quantum manifold spanned by two coherent states with opposite phase, which can be associated to the opposite magnetic states of a quantum $s=1/2$-spin \cite{Minganti2016,Bartolo2017}. This peculiar feature has motivated a deep research activity about these photonic systems, showing not only the feasibility of quantum computers and quantum annealers \cite{GotoSciRep2016,GotoPRA16,Nigg17,Puri2017,PuriNJP17}, but also the emergence of a second-order phase transition, analogous to that separating the paramagnetic and the ferromagnetic phases in quantum magnets \cite{Bartolo16,Savona17,Minganti18,Rota19}. In a rather different context, optical parametric amplifiers have been proposed as a coherent simulator of an Ising model \cite{Yamamoto2017}.

\begin{figure}
\begin{center}
\includegraphics[width=0.42\textwidth]{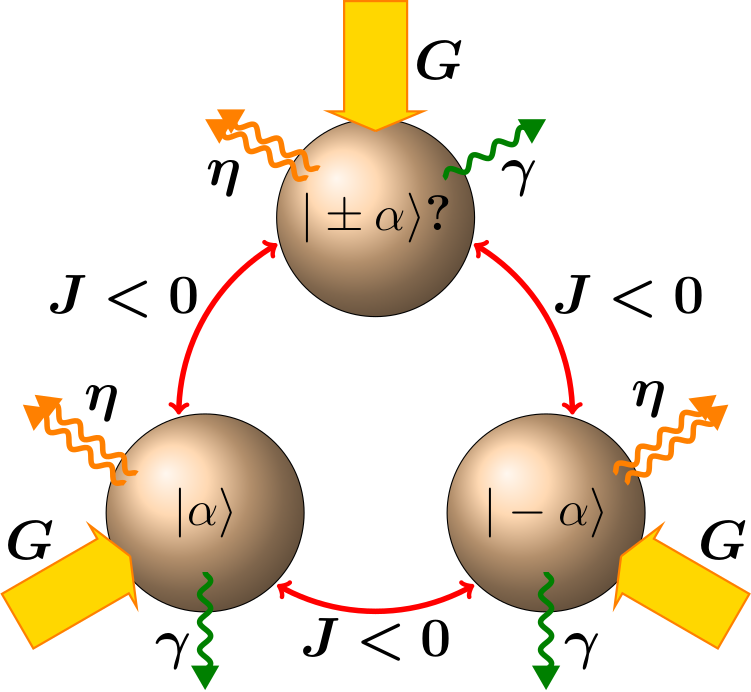}

\end{center}
\caption{A sketch of the array of cavities in presence of a two-photon driving with amplitude $G$, one-photon losses with rate $\gamma$, two-photon losses with rate $\eta$ and hopping amplitude $J<0$. In the limit of strong driving field, the photons in each cavity form a coherent state with phase $\alpha$ or $-\alpha$. The coupling with $J<0$ tends to favor the emergence of states with opposite phase between neighboring cavities, but this condition cannot be satisfied in the frustrated system made up of three mutually coupled cavities. This system bears strong analogies with three antiferromagnetically interacting Ising spins.}
\label{fig:sketch}
\end{figure}

In this work, we show how an array of coupled quadratically driven QED cavities can simulate the triangular antiferromagnetic Ising model \cite{Wannier50} \--- a well-known theoretical model supporting the emergence of a spin liquid phase. A necessary condition to recover this result is to engineer the coupling between the cavities such that the photon hopping strength is negative: this regime is experimentally feasible with photonic crystals \cite{Haddadi14} and a possible realization with QED cavities has been discussed recently \cite{Nigg17,Kounalakis2018}. By studying the first-order coherence correlation function, the entropy and the response to a single-photon driving field, we show that three coupled quadratically driven cavities, in the limit of strong two-photon pump, behave as three interacting spins with an Ising antiferromagnetic coupling (see Fig. \ref{fig:sketch}). 

The paper is organized as follows: in Sec. \ref{sec:methods} we describe the theoretical framework used in the work, including the definition of the relevant equations and the details of the numerical computations; in Sec. \ref{sec:results} we present and discuss the results obtained; finally, in Sec. \ref{sec:conclusion} we draw our conclusions.

\section{Theoretical model}\label{sec:methods}

We consider systems of $N = 2$ and $N = 3$ coupled photonic resonators, in presence of a Kerr nonlinearity with energy $U$ and two-photon driving with frequency $\omega_p$ and amplitude $G$. These can be modeled, in the reference frame rotating at half of the pump frequency, by the Hamiltonian (we set $\hbar = 1$) 
\begin{eqnarray}
\oH &=& \sum_{j=1}^N-\Delta\oad_j\oa_j + \frac{U}{2}\hat{a}^{\dagger  2}_j\oa_j^2 + \frac{G}{2} \hat{a}^{\dagger  2}_j+ \frac{G^*}{2} \oa_j^2\nonumber\\
&-& \sum_{j \ne j^\prime} \frac{J}{2} (\oad_j\oa_{j^\prime} + \oad_{j^\prime}\oa_{j}) \,.
\label{eq:hamiltonian}
\end{eqnarray}
where $\oa_j$ is the photon destruction operator acting on the $j$-th site. The quantity $\Delta = \omega_p/2 - \omega_c$ is the detuning between half of the two-photon driving field frequency $\omega_p$ and the resonant cavity frequency $\omega_c$. The photon hopping between different cavities, with strength $J$, is described by the last term in the equation.

Assuming Markovian dissipative processes for each cavity, the dynamics of the system is described by the density matrix $\orho(t)$ which obeys to the quantum master equation in the Lindblad form:
\begin{equation}
\frac{\partial \orho}{\partial t} = \mathcal{L} \orho = -i \left[\oH,\orho \right] + \sum_{j,k} \oGamma_{j,k}\orho\oGammad_{j,k}  - \frac{1}{2} \left\{\oGammad_{j,k}\oGamma_{j,k},\orho\right\} \ ,
\label{eq:lindblad-me}
\end{equation}
where $\mathcal{L}$ is the Liouvillian superoperator and the jump operators $\oGamma_{j,k}$ describe the transition induced by the environment on the system. In the case of quadratically driven cavities, it is necessary to consider two different kinds of dissipative processes. First, one-photon losses, modeled by the jump operators $\oGamma_{j,1} = \sqrt{\gamma}\oa_j$. Furthermore, since we assume an input channel injecting photons in pairs, then dissipative processes will likely arise through the same channel and therefore it is necessary to consider two-photon losses, which are described by the jump operators $\oGamma_{j,2} = \sqrt{\eta}\oa_j^2$.

The dynamics of the system evolves at large times towards a steady-state $\orho_{ss}$, which satisfies the condition $\partial \orho_{ss}/\partial t = 0$. We determine the steady-state density matrix by numerically solving the linear system $\mathcal{L} \orho_{ss} = 0$, with the constraint $\textrm{Tr}(\orho_{ss}) = 1$. The Hilbert space is truncated by setting a maximum value $N_m$ for the photon occupancy per cavity and a maximum value $N_{m,T}$ for the total photon occupancy in the system: the accuracy of the numerical results is checked by studying their convergence with $N_m$ and $N_{m,T}$. The optimal values of $N_m$ and $N_{m,T}$ depend on the particular physical parameters of each simulation. For the regimes where photon occupation is the largest (i.e. large $G/\gamma$ and small $U/\gamma$), for the case of $N=3$ cavities, the convergence is reached for $N_m = 22$ and $N_{m,t} = 42$: this corresponds to a Hilbert space of $\sim 10^4$ and hence a master equation for the steady-state density matrix (Eq. \eqref{eq:lindblad-me}) equivalent to a linear system of $10^8$ equations.

In Ref. \cite{Rota19}, it has been shown that a system of coupled quadratically driven cavities can be approximated by a spin-$1/2$ lattice governed by a $XY$ Hamiltonian in a transverse field:
	\begin{equation}
	\oH_{XY} = h_z \sum_j \hat{\sigma}_{j}^{(z)} - J_{XY} \sum_{\langle j,j^\prime\rangle} \left[\eta_x \hat{\sigma}_{j}^{(x)} \hat{\sigma}_{j^\prime}^{(x)} + \eta_y \hat{\sigma}_{j}^{(y)} \hat{\sigma}_{j^\prime}^{(y)}\right] \ . \label{eq:hamiltonianXY} 
	\end{equation}
	The coupling strength $J_{XY}$ in the effective model is proportional to the photon hopping strength $J$ of the bosonic system. The case of $J>0$ -- studied in Ref. \cite{Savona17} within a mean-field approximation and in Ref. \cite{Rota19} using a fully many-body approach -- shows that the quadratically-driven Bose-Hubbard model presents a steady state that is a statistical mixture of two equiprobable separable coherent states $\ket{\Psi_\pm} = \prod_{j} \ket{\pm \alpha}_j$ in the limit of large $G$. These states can be associated to a ferromagnetic phase, if one associates the local coherent states with opposite $\alpha$ to the spin states with opposite magnetization. The case of $J<0$ should then simulate an antiferromagnetic coupling in the approximate spin model, which corresponds in an extended 1D arrays of photonic cavities to the emergence of states $\ket{\Phi_{\pm}} = \ket{\pm \alpha, \mp \alpha, \pm \alpha, \mp \alpha, \ldots}$ with $k = \pi$ modulation.

In order to investigate this simulated antiferromagnetic spin model, we study the steady-state properties of the photonic system varying the value of the two-photon driving amplitude $G$ (which we always consider real-valued) and assuming $\Delta = J < 0$ in Eq. \eqref{eq:hamiltonian}. This latter condition corresponds to setting the two-photon driving field resonant with the $k=\pi$-mode of the single particle spectrum of the Bose-Hubbard Hamiltonian. This choice should thus favor the emergence of states $\ket{\Phi_{\pm}}$ in the steady state of the open system. We point out that, for arrays of finite length with periodic boundary conditions, the $k=\pi$-mode is present in the energy spectrum of the closed system only if the number of sites is even. For an odd number of sites, the geometrical frustration makes the two-photon pump off-resonant with any of the eigenstates of the Bose-Hubbard Hamiltonian. Hence, it is expected that signatures of an antiferromagnetic behavior emerge at larger values of the driving field amplitude in a frustrated array, than in a commensurate one.


\section{Results}\label{sec:results}

To show the analogies between our quadratically driven photonic system and a frustrated antiferromagnet, we focus at first on the 
first-order coherence correlation function
\begin{equation}\label{eq:DefCorrelation}
g^{(1)}_{1,2} = \frac{\textrm{Tr}\left( \orho_{ss} \oa_1^\dagger \oa_2 \right)}{\textrm{Tr}\left( \orho_{ss} \oa_1^\dagger \oa_1 \right)} \ , 
\end{equation}
and the von-Neumann entropy
\begin{equation}\label{eq:DefEntropy}
S = - \textrm{Tr} \left(\orho_{ss} \ln \orho_{ss} \right) \ .
\end{equation}

In particular, $g^{(1)}_{1,2}$ is related to a spin-spin correlation function. Indeed, if we adopt the approximate mapping of the photon annihilation operator onto spin operators of the effective model \cite{Rota19} to Eq. \eqref{eq:DefCorrelation}, we obtain
\begin{equation}
g^{(1)}_{1,2} \simeq \frac{B_+^2  \average{\hat{\sigma}^{(x)}_1 \hat{\sigma}^{(x)}_2} + B_-^2  \average{\hat{\sigma}^{(y)}_1 \hat{\sigma}^{(y)}_2}}{(B_+^2 + B_-^2) + 2 B_+ B_- \average{\hat{\sigma}^{(z)}_1}} \,.
\end{equation}
Here, $B_{\pm} = \sqrt{\tanh(|\alpha|^2)} \pm (\sqrt{\tanh(|\alpha|^2)})^{-1}$ and $\alpha$ is a parameter determined uniquely by the values of the system parameters, which can be interpreted as the field amplitude of an optimal local coherent state. For $G/\gamma \to \infty$, one has $|\alpha|\to \infty$ and $B_- \to 0$, leading to $g^{(1)}_{1,2} \simeq \average{\hat{\sigma}^{(x)}_1 \hat{\sigma}^{(x)}_2}$.

\begin{figure}
	\includegraphics[width=0.48\textwidth]{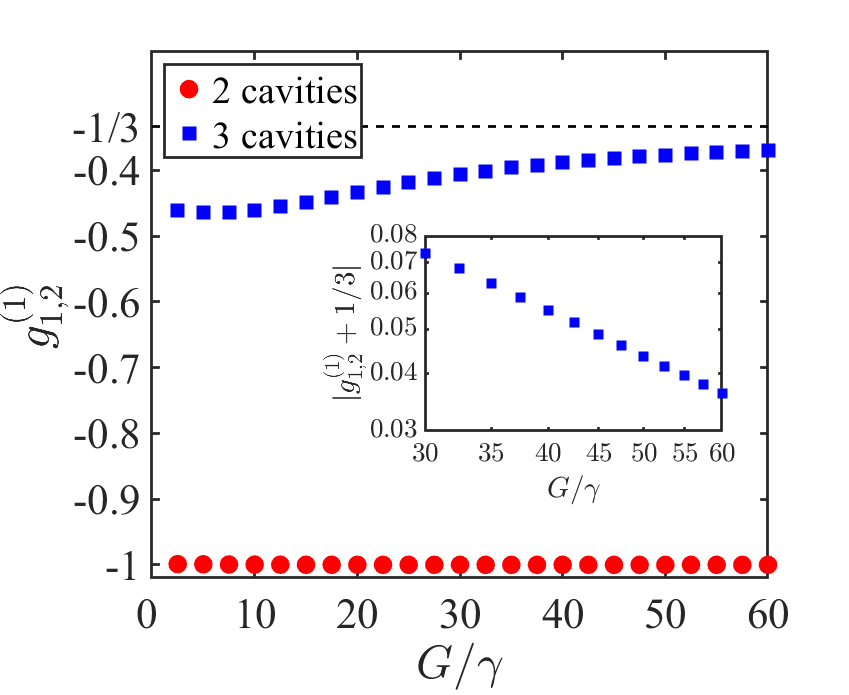}
	\caption{The first-order correlation function $g^{(1)}_{1,2}$ as a function of the amplitude of the two-photon driving, for the systems of $N = 2$ and $N = 3$ cavities. Inset: behavior of $g^{(1)}_{1,2}$ vs. $G$ in the regime of large $G/\gamma$ for the case of $N = 3$ cavities, plotted on a log-log scale: the results show a power law dependence $|1/3+g^{(1)}_{1,2}| \sim G^\nu$, with $\nu = -1.02 \pm 0.02$. The other Hamiltonian parameters are $U/\gamma = 10$, $\Delta/\gamma = J/\gamma = -10$.}\label{fig:correlation}
\end{figure}

\begin{figure}
	\includegraphics[width=0.48\textwidth]{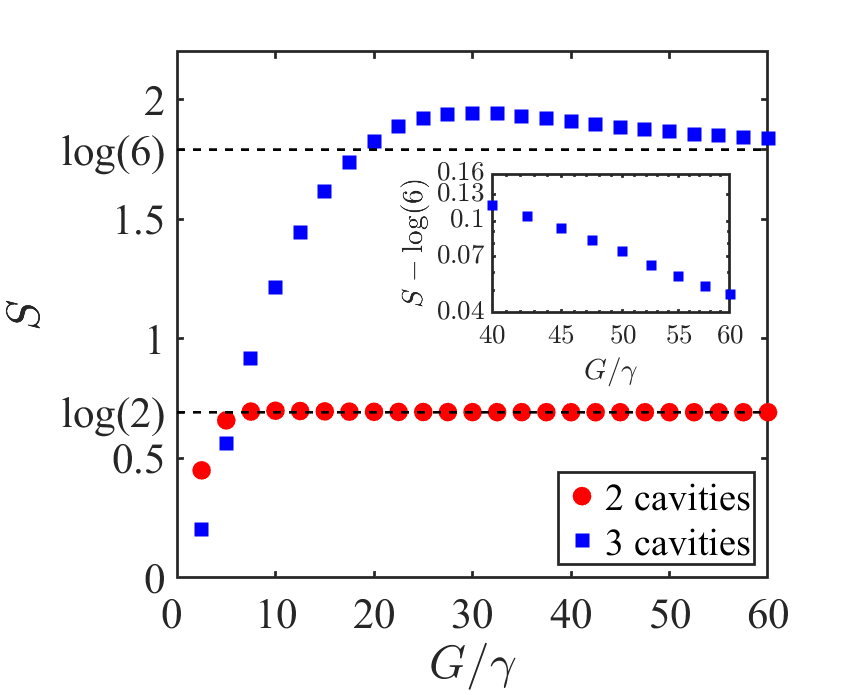}
	\caption{The von-Neumann entropy $S$ as a function of the amplitude of the two-photon driving, for the systems of $N = 2$ and $N = 3$ cavities. Inset: behavior of $S$ vs. $G$ in the regime of large $G/\gamma$ for the case of $N = 3$ cavities, plotted on a log-log scale: the results show a power law dependence $S - \ln(6) \sim G^\mu$, with $\mu = -2.28 \pm 0.08$. The other Hamiltonian parameters are $U/\gamma = 10$, $\Delta/\gamma = J/\gamma = -10$.}\label{fig:entropy}
\end{figure}


The behavior of $g^{(1)}_{1,2}$ and of $S$ as a function of the driving field amplitude $G$ are shown respectively in Fig. \ref{fig:correlation} and \ref{fig:entropy}, both for the system with $N=2$ and $N=3$ cavities. The results for the system made up of $N=2$ cavities bear a clear signature of an antiferromagnetic interaction. The correlation function $g^{(1)}_{1,2}$ is negative and, for increasing $G$, approaches the asymptotic value $g^{(1)}_{1,2} = -1$. Moreover, the entropy as a function of $G$ increases from the value $S = 0$ in the limit of a vanishing driving amplitude (notice that for $G = 0$, the steady state is pure and corresponds to the bosonic vacuum) to the asymptotic value $S = \ln(2)$ in the limit $G/\gamma \to \infty$, indicating that the steady-state density matrix is dominated by two equiprobable eigenstates.

The results for $g^{(1)}_{1,2}$ and for $S$ indicate that, in the limit of $G\gg\gamma$, the steady state is described by a statistical mixture of two separable states, obtained as product of two local coherent states with opposite phase. Its density matrix can therefore be written as
\begin{equation}\label{eq:rhoApprox2}
\orho_2 = \frac{1}{2} \ket{\alpha_0,-\alpha_0}\bra{\alpha_0,-\alpha_0} + \frac{1}{2} \ket{-\alpha_0,\alpha_0}\bra{-\alpha_0,\alpha_0} \ .
\end{equation}

To test this assumption, we compute the fidelity $\mathcal{F}(\orho_2,\orho_{ss}) = \left( \textrm{Tr} \left(\sqrt{\sqrt{\orho_2} \orho_{ss} \sqrt{\orho_2}}\right) \right)^2$ the steady-state density matrix $\orho_{ss}$ obtained by the numerical solution of the master equation \eqref{eq:lindblad-me} and the density matrix $\hat{\rho}_2$ in Eq. \eqref{eq:rhoApprox2}, with the phase $\alpha_0$ of the local coherent states obtained as $\alpha_0 = \sqrt{\textrm{Tr}(\orho_{ss} \oa_1^2)}$. It turns out that $1 - \mathcal{F}(\orho_2,\orho_{ss}) < 10^{-4}$ for all the values $G /\gamma \ge 30$, thus indicating that Eq. \eqref{eq:rhoApprox2} is a very good description of the steady-state density matrix in the limit of strong driving.


The possibility to investigate the effects of frustration is highlighted in the results of the system made up of $N = 3$ cavities. In this case, the coupling between different cavities is at odds with the geometric constraints of the system, thus leading to a behavior similar to frustrated antiferromagnets. The correlation function $g^{(1)}_{1,2}$ (Fig. \ref{fig:correlation}) presents a non monotonous behavior as a function of $G$ and, for large values of $G/\gamma$, it converges with a power law behavior to the asymptotic value $-1/3$. This value is typical of the spin-spin correlation function between nearest neighbors in a triangular antiferromagnetic Ising model \cite{Stephenson64}. The behavior of the entropy $S$ (Fig. \ref{fig:entropy}) also confirms the analogy with an antiferromagnetic system. In this case, the frustration in the spin alignment results in the appearance of six possible configuration minimizing the energy (i.e. all those with two antiparallel and one parallel pair of spin). The asymptotic value reached in the limit of large $G/\gamma$ is $S = \ln(6)$, consistently with the six-fold degeneracy of the ground state of the equivalent spin model.  

In analogy to the case with $N=2$, we can construct an approximation for the steady-state density matrix for the system of $N=3$ cavities in the limit where $G\gg\gamma$ as a statistical mixture of states obtained as tensor products of local coherent states with opposite phase. From the analogy with the ground state of the three antiferromagnetically coupled spins, the approximate steady state is
\begin{eqnarray}\label{eq:rhoApprox3}
\orho_3 = \frac{1}{6} & & \left( \ket{\alpha_0,\alpha_0,-\alpha_0}\bra{\alpha_0,\alpha_0,-\alpha_0} + \right. \nonumber \\
& & \ket{\alpha_0,-\alpha_0,\alpha_0}\bra{\alpha_0,-\alpha_0,\alpha_0} + \nonumber \\
& & \ket{-\alpha_0,\alpha_0,\alpha_0}\bra{-\alpha_0,\alpha_0,\alpha_0} + \nonumber \\
& & \ket{-\alpha_0,-\alpha_0,\alpha_0}\bra{-\alpha_0,-\alpha_0,\alpha_0} + \nonumber \\
& & \ket{-\alpha_0,\alpha_0,-\alpha_0}\bra{-\alpha_0,\alpha_0,-\alpha_0} + \nonumber \\
& & \left. \ket{\alpha_0,-\alpha_0,-\alpha_0}\bra{\alpha_0,-\alpha_0,-\alpha_0}  \right) \ . \label{eq:rhoApprox3}
\end{eqnarray}

At $G/\gamma = 60$, we have $\alpha_0 = 0.112 - 2.299i$ and the fidelity between the steady-state density matrix $\orho_{ss}$ and the approximation given by Eq. \eqref{eq:rhoApprox3} is $\mathcal{F}(\orho_3,\orho_{ss}) = 0.956$.

	\begin{figure}
		\includegraphics[width=0.48\textwidth]{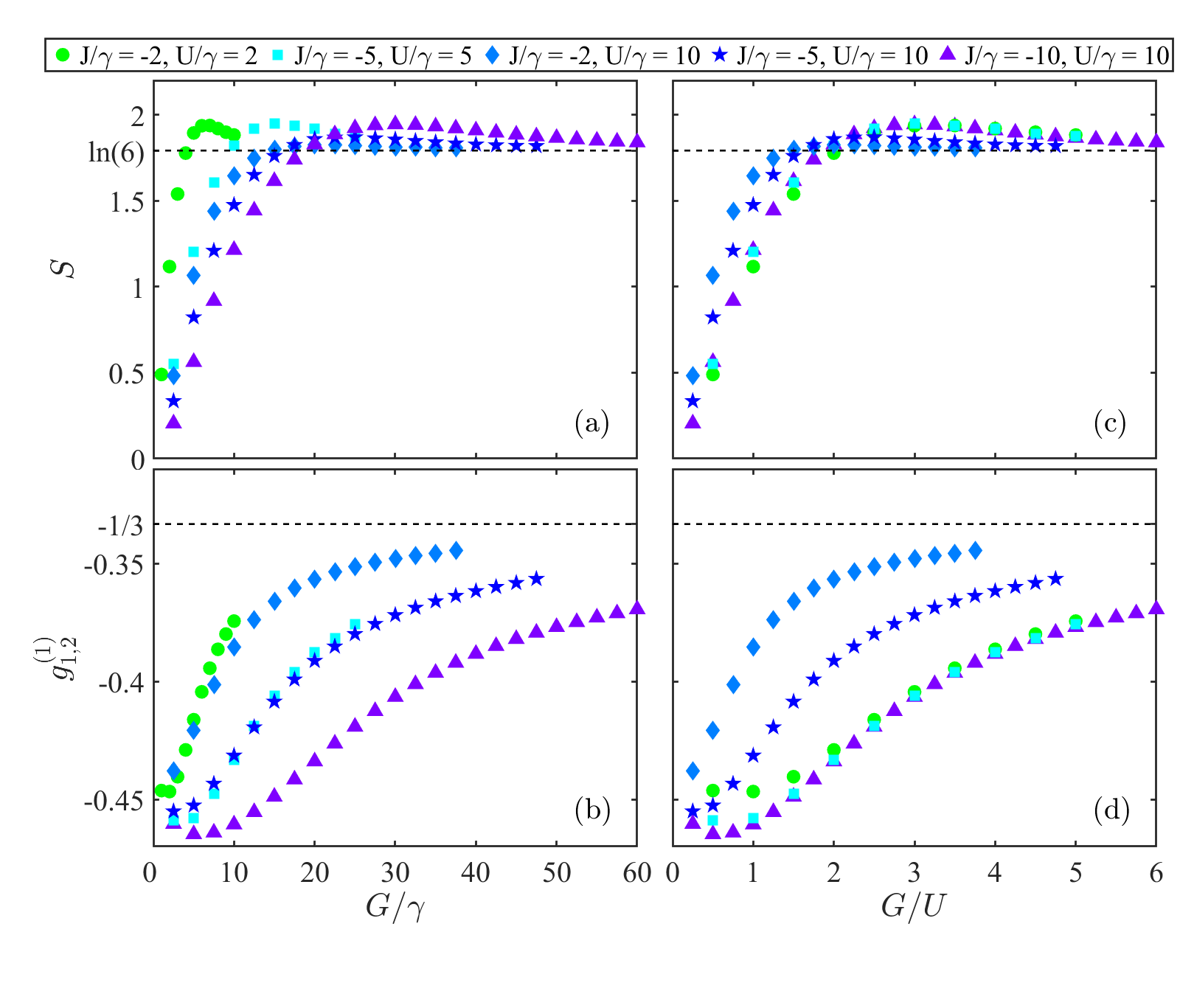}
		\caption{The von-Neumann entropy $S$ (panel a) and the first-order coherence correlation function $g^{(1)}_{1,2}$ (panel b) as a function of the amplitude of the two-photon driving rescaled with the one-photon loss rate $G/\gamma$ for the system of $N = 3$ cavities and for different values of the Hamiltonian parameters. The same data are plotted in panels (c) and (d) as a function of the two-photon driving rescaled with the non-linearity $G/U$.}\label{fig:robustness}
	\end{figure}
In Fig. \ref{fig:robustness}, we show the results for the von-Neumann entropy $S$ and the correlation function $g^{(1)}_{1,2}$ of the system of $N=3$ cavities for different values of the Hamiltonian parameters. Although the different parameters affect quantitatively the curves of $S$ and $g^{(1)}_{1,2}$ as a function of $G/\gamma$ (panels (a) and (b)), their qualitative behavior is independent of $J$ and $U$. Indeed, the results in Fig. \ref{fig:robustness} indicates that, for large enough two-photon driving, $S$ and $g^{(1)}_{1,2}$ tend to the corresponding limiting values $\ln(6)$ and $-1/3$ for all the values of $J$ and $U$. Interestingly, when the data are shown as a function of the ratio $G/U$ (panels (c) and (d)), the behavior of $S$ and $g^{(1)}_{1,2}$ depends on the value of $J/U$ but is independent of the value $\gamma/U$. This last analysis indicates that the simulated antiferromagnetic behavior is robust with respect to dissipation and does not require low-loss cavities to be observed. Moreover, it supports the statement that the antiferromagnetic behavior of the photonic simulator is related exclusively to the local $\mathbb{Z}_2$ symmetry, rather than to a particular regime of physical parameters.

		\begin{figure}
			\includegraphics[width=0.48\textwidth]{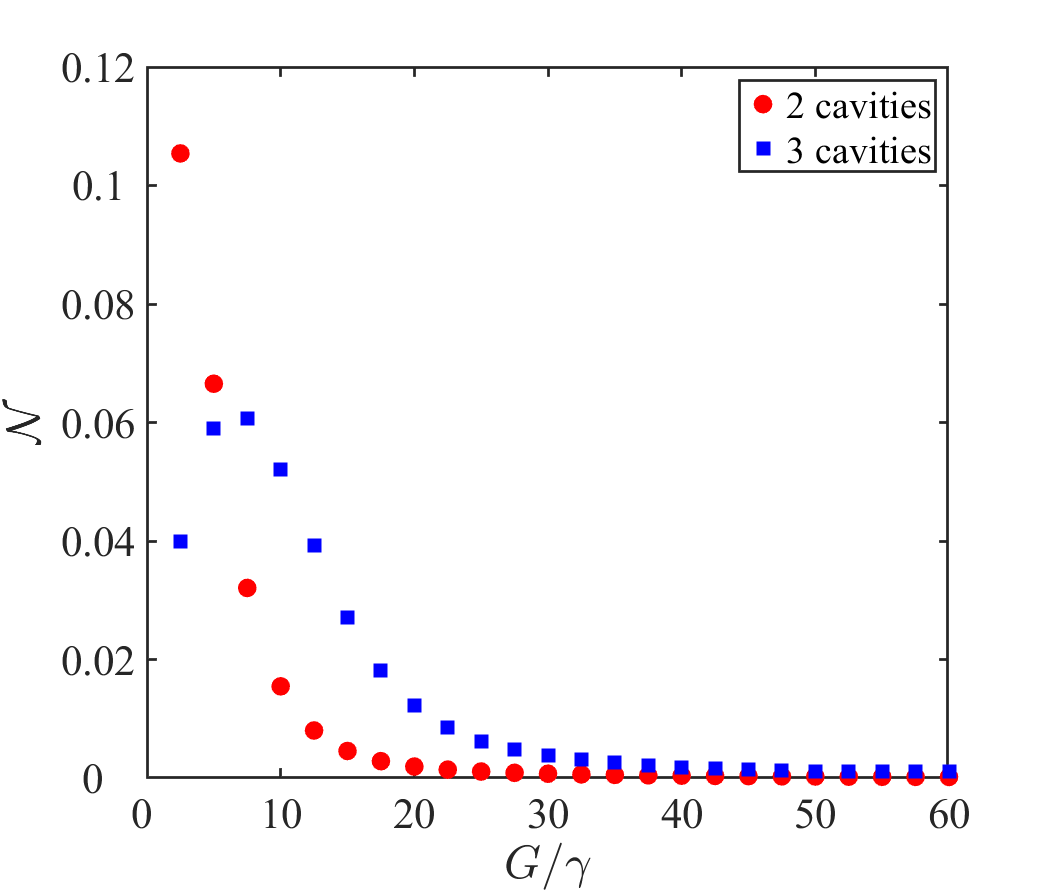}
			\caption{The entanglement negativity $\mathcal{N}$ as a function of the amplitude of the two-photon driving, for the systems of $N = 2$ and $N = 3$ cavities. The other Hamiltonian parameters are $U/\gamma = 10$, $\Delta/\gamma = J/\gamma = -10$.}\label{fig:negativity}
		\end{figure}
	It is important to notice that, even though the steady state of the system is separable in the limit of strong two-photon driving, our simulator can support fully quantum correlated states. This can be deduced from the computation of the entanglement negativity $\mathcal{N}$, from the partial transpose of the steady-state density matrix with respect to one of the $N$ cavities in the array. The results for $\mathcal{N}$ are presented in Fig. \ref{fig:negativity} and show clearly the presence of entanglement ($\mathcal{N} > 0$ for all $G/\gamma$), both in the commensurate and frustrated lattice. Only in the limit of $G/\gamma \to \infty$ the negativity goes to zero, as it is expected for the classical steady states in Eq. \eqref{eq:rhoApprox2} and \eqref{eq:rhoApprox3}.

\begin{figure}
	\includegraphics[width=0.48\textwidth]{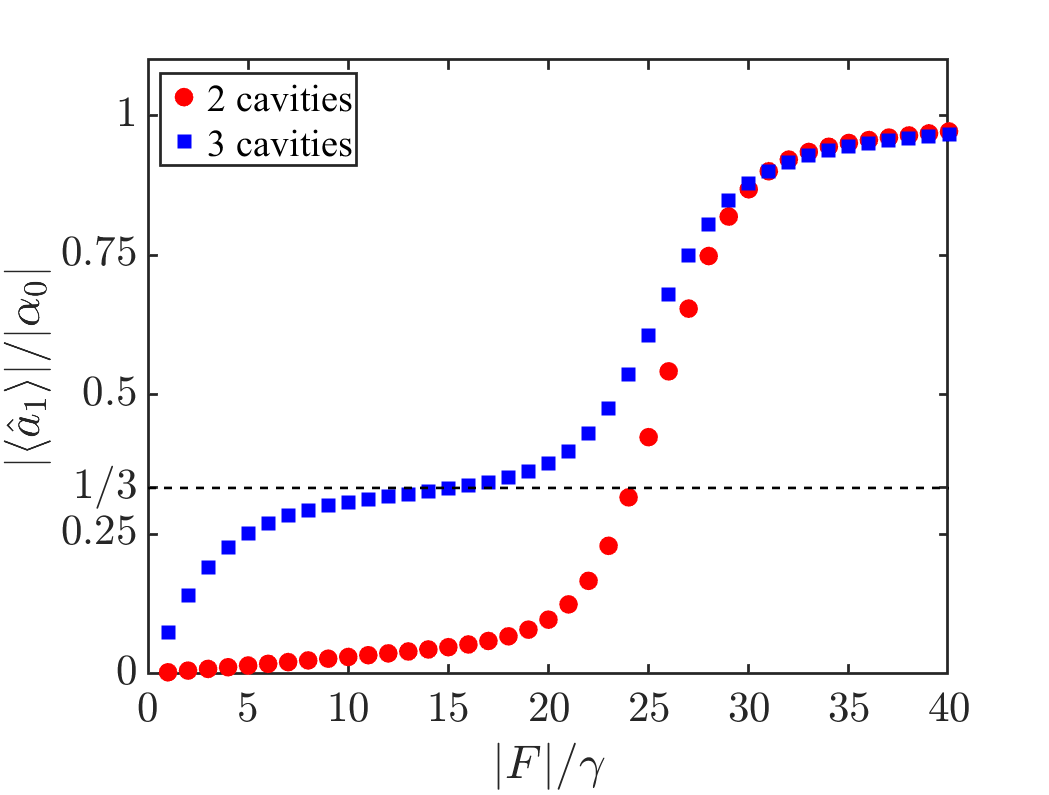}
	\caption{Field amplitude $|\average{\oa_1}|$ (rescaled by $|\alpha_0|$) as a function of the amplitude $F$ of the applied one-photon driving, for the systems of $N = 2$ and $N = 3$ cavities. The other parameters of the Hamiltonian are  $U/\gamma = 10$, $\Delta/\gamma = J/\gamma = -10$, $G/\gamma = 60$}\label{fig:response}
\end{figure}

A further evidence of the spin analogy can be found in the non-linear response of the quadratically driven photonic system to an one-photon pump. This latter can be modeled with an additional term in the Hamiltonian of Eq. \ref{eq:hamiltonian}:
\begin{equation}\label{eq:HamiltonianwithField}
\oH_{F} = \oH + \sum_{j}(F \oa^\dagger_j + F^* \oa_j) \ .
\end{equation}
According to the spin approximation discussed in Ref. \cite{Rota19}, this term can be associated to an external field in the magnetic analog of our system. The direction of the analog external field depends on $\alpha_0$ and, in the limit of large $\alpha_0$ (i.e. large $G/\gamma$), it becomes parallel to the direction of the Ising antiferromagnetic coupling. We have calculated the steady-state density matrix $\orho_{F}$ of the system in presence of a strong two-photon driving $G/\gamma = 60$ and a variable one-photon driving $F$. We show in Fig. \ref{fig:response} the expectation value of the induced coherence $\average{\oa_1} = \textrm{Tr}(\orho_{F} \oa_1)$. The quantity $\average{\oa_1}$ strongly depends on the phase of the one-photon driving $F$, and the effect of the one-photon pump is more evident when the quantity $F^* \alpha_0$ is purely real. For this reason, in the results of Fig. \ref{fig:response}, we have set the phase of $F$ according to this condition and vary the absolute value $|F|$.

The dependence of $\average{\oa_1}$ on $|F|$ is particularly different in the two systems with $N=2$ and $N = 3$ cavities. For $N=2$, we can distinguish two different regimes, according to the amplitude of the one-photon driving. For small $|F|$ (up to $|F|/\gamma \lesssim 20$), the induced coherence increases slowly and linearly with the pump, indicating that the system maintains a certain antiferromagnetic order, as it happens in the absence of the one-photon driving. For large $|F|$, instead, the antiferromagnetic order is broken and the steady state of the photonic system is a pure coherent state, with $\average{\oa_1} = \alpha_0$. For the system with $N=3$ cavities, the behavior of $\average{\oa_1}$ is similar to the previous case only for large $|F|$, but is notably different in the opposite regime. For small $|F|$, we notice that, after a steep increase of $\average{\oa_1}$ with $|F|$ at very small values of the pump, there is a a broad interval where the induced coherence depends very weakly on $|F|$ and stabilizes around a value close to $\alpha_0/3$. This latter behavior is reminiscent of the $1/3$-magnetization plateau which emerges in the triangular antiferromagnetic Ising model in presence of an external magnetic field along the direction of the coupling \cite{Metcalf1973}. It corresponds to the minimal energy configurations where two thirds of the spins in the lattice point in the direction of the external field and the remaining one third in the opposite one.


\section{Conclusions}\label{sec:conclusion}

In conclusion, we have considered a system of coupled photonic cavities subject to a two-photon driving and showed the existence of regimes where these can simulate the properties of Ising antiferromagnets. The key feature, allowing the emergence of antiferromagnetic correlations among the photonic states, is a negative hopping rate between different cavities, a condition already realized experimentally \cite{Haddadi14}. By comparing the behavior of the systems of two and three cavities, whose geometry are respectively commensurate and incommensurate to the antiferromagnetic coupling, we highlighted the effects due to the frustration of the lattice, analogous to those arising in spin models. The von-Neumann entropy in particular signals the increased fluctuations in the frustrated system, which can be ascribed to a larger degeneracy of the states at minimum energy. The response of the photonic system to a coherent one-photon drive shows the emergence of a plateau in the induced coherence, which is reminiscent of the behavior of frustrated antiferromagnets under an external magnetic field.

Thanks to the possibility of realizing and manipulating systems of quadratically driven nonlinear photonic cavities within current experimental techniques, our results point to a novel class of quantum simulators for antiferromagnets, which could allow to investigate the properties of spin liquids by means of a fully controllable and versatile experimental platform. From the theoretical point of view, an important question which should be addressed in the future concerns the possibility to achieve strong quantum correlations in this system, and the effects the geometric frustration can have on these. This possibility could be relevant to investigate the elusive quantum spin liquid phase.


We acknowledge enlightening discussions with Cristiano Ciuti and Fabrizio Minganti.

\bibliographystyle{apsrev4-1}
\bibliography{CavitiesFrustrationBIB}

\end{document}